\documentclass{optica-article}

\journal{opticajournal} % for journals or Optica Open

\articletype{Research Article}
 \usepackage{lineno}
 \usepackage{float}
 \usepackage[T1]{fontenc}
 \usepackage[utf8]{inputenc}

\begin{document}

\title{Toward high-speed effective numerical simulation of multiple filamentation of high-power femtosecond laser radiation in transparent medium}  

\author{Andrey Bulygin,\authormark{1,2}  and Yury Geints\authormark{1}}

\address{\authormark{1} V.E. Zuev Institute of Atmospheric Optics SB RAS, Tomsk, Russia\\
\authormark{2}Tomsk State University, Lenin ave. 36, Tomsk
634050, Russia\\
}

\email{\authormark{*}bulyginandrey7@gmail.com} %% email address is required; see note below about the corresponding author designation

% use {asbstract*} to suppress the copyright line. Copyright information will be added in production

\begin{abstract*} 
High-power femtosecond laser radiation during the propagation in air (and other transparent media) experiences multiple filamentation. Filamentation is a unique nonlinear optical phenomenon, which is accompanied by a wealth of nonlinear optical effects such as formation of extended plasma channels in the beam wake, generation of higher harmonics and supercontinuum, generation of THz radiation. The manifestations of laser filamentation can be useful for solving atmospheric optics problems related to remote sensing of the environment as well as directed transmission of laser power. The classical numerical methods used for simulating the nonlinear long-range atmospheric propagation of high-power radiation with a sufficiently large laser beam aperture have almost reached their limit regarding the acceleration of calculations. To solve this problem and speed-up the numerical simulations of laser filamentation, we propose an improved numerical technique based on a modified method of phase screens constructed on a sparse spatial grid. Within the framework of this technique, we seek for optimal ansatz (substitution function) to the governing equations using the machine learning technology, which provides for the best correspondence to the numerical solution of the test problem using a denser spatial grid.

\end{abstract*}
\section{\label{sec:level1} Introduction }
The propagation of high-power femtosecond laser radiation (HPLR)  in air usually occurs in a nonlinear regime. In media with optical nonlinearity of the cubic type (Kerr-type nonlinearity), which include most gases and transparent dielectrics, the self-focusing of optical radiation leads to a bright nonlinear phenomenon called laser filamentation\cite{Boyd,COUAIRON_2007,Chekalin:2013}.   Pulse filamentation is accompanied by strong aberrations of the optical radiation, namely the large-scale transformations of its spatial-temporal profile. This leads to the formation of localized high-intensity optical structures usually referred to as the filaments. A characteristic feature of laser filaments is their stable transverse size, which persists over a sufficiently long distance and can be longer than the Rayleigh length of a beam of this diameter. In atmospheric air, the peak intensity in the filament can reach several hundreds of $TW/cm^2$, while the average filament size varies from several tens to hundreds of micrometers depending on air pressure and laser wavelength  \cite{Champeaux:06,Geints:14,Mitrofanov:18}.
Experimentally, the laser pulse filamentation usually manifests itself as the appearance of glowing filaments in the visible spectrum in the beam channel. This is due to the recom-bination of plasma electrons in the regions formed as a result of laser pulse induced air molecules ionization (tunnel, multi-photon). The numerical simulation of the nonlinear propagation in air of a small-aperture (diameter of several millimeters or less) femtosecond laser pulse is usually carried out within the nonlinear Schrodinger equation (NLSE) in (2D+1) or (3D+1) formulation or using the more rigorous analog known as the unidirectional pulse propagation equation (UPPE) [2]. Note, that this is not an easy computational task, since it requires a sufficiently dense computational grid with a transverse spatial step of about tens of micrometers and a longitudinal step less than a few millimeters. This significantly complicates the implementation of massive in-parallel numerical calculations and limits the maximum range of pulse propagation simulation to several tens of meters \cite{Voronin:2016,HU2017281}.
At the same time, the practical needs of nonlinear atmospheric optics demand at least kilometer-long optical distances upon taking into account various optical weather (atmos-pheric aerosol, turbulence). Such long-range nonlinear propagation distances can be achieved only with wide centimeter-scale femtosecond laser beams, because of the fundamental limitation of the nonlinear focus position (the filamentation region) imposed by the Rayleigh diffraction length. When transferring to centimeter-diameter laser beams, the amount of computer resources and time required for numerical simulation of even one implementation of the laser pulse propagation increases enormously. It can be argued that at the present stage of computer technology development, direct numerical calculation of laser filamentation for such experiments within the complete non-stationary model is impractical even with the use of the supercomputer clusters. Therefore, to simplify the numerical calculation, as a rule, a reduction in the problem spatial dimensions is used by moving to its stationary approximation in various forms \cite{Dergachev_2014,Stotts:22,Balashov_2006,Rytov_2008} .
For example, in \cite{Fedorov}, a phase-modulated (chirped) pulse was simulated through the substitution of real pulse power dependence on the path length by a simple analytical relationship determined by the dispersion law. However even in such case, one can only talk about the possibility of predicting the filamentation onset, and not about investigation its dynamics. Several efforts also have been made to find the analytical solutions and to study the properties of the NLSE in case of laser filamentation. For example, in \cite{PhysRevLett.114.063903} it is theoretically shown that multiple pulse filamentation can take place in different phase modes (phase transitions of the second kind), which correlates with the experimental results \cite{Petrarca}.
This finding is qualitatively consistent with that in Ref. [16], where based on the self-consistent field approximation the authors propose to restore the dynamics of the HPLR propagation as a whole, i.e., on the macroscopic scale (scale much larger than the size of a single filament). However, for practical implementation of these ideas there is not enough validity and verification on a set of test cases.
It should be noted, that along with numerical approaches to the problem of laser fila-mentation there are also attempts to obtain some analytical solutions to the problem con-sidered, in particular, using the geometric optics (GO) approach \cite{Garcia,Vuong,Semak}. However, the use of GO approximation can be justified only in the pre-filamentation region. Worthwhile noting, the transition from studying the small-scale optical field dynamics toward the pulse integral parameters, such as the total pulse energy and the root-mean-square (effective) beam radius. Here, one can mention the works \cite{Bulygin_13,Bulygin_18,Couairon}, which analyze the evolution of the effective radius and other integral properties of laser radiation during the filamentation in air including the use of completely conservative numerical schemes. The knowledge about the behavior of the effective beam radius can be useful both in practical atmospheric optics applications and in the theory of pulse self-focusing as it gives important qualitative ideas about the physics of filamentation. If one considers the actual numerical methods for solving the problem of nonlinear focusing of a high-power laser beam in the atmosphere, then in addition to semi-analytical approaches the attempts are made to develop various simplified methods.
Here, the method of averaged optical field of a laser pulse is worth mentioning \cite{Bulygin_13}. Despite certain theoretical successes in this field, which led to the theoretical discovery of the filamentation mode change as a second-order phase transition, this method has significant drawbacks. They stem from the fact that the transition to a macroscopic description of field dynamics contains many heuristic constructions, and this makes this averaged method ill-conditioned regarding the evidence base. Additionally, the description of the transient regime of pulse self-action is missed, when direct numerical calculation is already difficult and the use of a macroscopic description is yet incorrect. Obviously, a certain transitional bridge from the NLSE-based description to the macroscopic approach should be developed, which will allow not only to verify this approach but also to substantiate and validate it on an array of numerical calculations.
In this paper, we propose a novel theoretical approach which is based on the classical method of phase screens to improve the efficiency of the numerical simulations of the problem on the long-range nonlinear propagation of a wide-aperture HPLR in a nonlinear randomly inhomogeneous medium (atmosphere) under conditions of laser pulse self-focusing and multiple filamentation. We demonstrate a new approach to dramatically accelerate the numerical simulation of a wide-aperture HPLR propagation in the real atmosphere.
The idea of the technique proposed is to replace the region of active optical radiation interaction with nonlinear inhomogeneous medium (filamentation region, aerosol cloud, turbulent layer) with certain effective phase screen(s) given in the form of an Ansatz, the parameters of which are found by means of the machine learning (ML) methods based on the condition of the closest match to the test problem solutions on the numerical grids with sufficiently dense nodes. Importantly, the simplest situations including narrow laser beams filamentation, are directly calculated with required accuracy and used as a database of the test solutions, as well as the available experimental data on HPLR propagation on real atmospheric paths are also used to obtain a representative array of solutions of the HPLR propagation problem.
Worthwhile, the classical numerical methods used to solve this problem have almost reached the limit in accelerating calculations when simulating the long-range atmospheric HPLR propagation with a sufficiently large beam aperture (centimeters). Nowadays, the ef-ficiency increasing of all known numerical methods is possible only by low-level optimization of the software implementations. The semi-analytical method laid down in this work does not cancel the existing developments but completes them with new methods including ML practices.

\section{\label{sec:level2} Math and Equations } 
The Nonlinear Schrodinger equation for HPLR propagation in a nonlinear medium (air) can be considered in general form:
\begin{equation}\label{eq1} 
2i k_0 \partial_z U =(\Delta_{\perp} + \epsilon_{k} UU^* +\epsilon_{h}[U] ) U  \equiv (\hat{h}_{k}+\epsilon_{h}[U])  U
\end{equation}
%2ik_0 U_z=(Δ_⊥+ϵ_ker (UU^*)+ϵ_eff [U])U				(1)
Here, $k_0$ is the wave number at the carrier wavelength (800 nm),$\epsilon_{k}$ is the coefficient for cubic medium nonlinearity (Kerr nonlinearity), $\epsilon_{h}$ is some (yet formal) complex medium dielectric permittivity associated with the manifestation of higher-order optical nonlinearities.
Obviously, when solving Eq. (1) analytically or numerically the main difficulty is the spatial region, where $\epsilon_{h} \ne 0$. Indeed, this is the pulse filamentation region or in a more general formulation the region of strong plasma nonlinearity. Within the framework of the split-step method according to different physical mechanisms, the propagation of a laser radiation through the entire filamentation region we replace by the scattering of a laser pulse at a certain spatially lumped complex effective phase screen (ES).

Meanwhile, one strives to ensure that the scattering at this ES gives the effect as close as possible to that which is realized in the classical filamentation model formulated on a denser numerical grid.
This approximation can be corrected in such a form if the region of higher nonlinearities has strongly pronounced properties of spatial localization or in other words, discreteness. This is exactly the peculiarity of the model considered below.  And then if the characteristic longitudinal scale of manifestation  $\epsilon_{h}$ is $l_{f}$ then formally, the optical radiation scattering at an effective screen can be represented in the following operator form:
\begin{eqnarray}\label{eq2} 
{U_{out}}=&&e^{-i\int_{z}^{z+l_{f}} ((\hat{h}_{k}+\epsilon_{h}))/(2 k_0) dz }  U_{in}\nonumber\\ 
&&\equiv
e^{-i(\hat{h}_{k}l_f/(2 k_0)-\hat{D}_f [U_{in}])  } U_{in}\nonumber\\ 
&&\approx
e^{-i(\hat{h}_{k}l_f/(2 k_0)-f_{lens} [U_{in},D])  } U_{in}
\end{eqnarray}

Here, $U_{in}$ is the optical field before the ES, while $U_out$ is the field that is formed as a result of the scattering on the ES, which is denoted as operater $\hat{D}_f [U_{in}]$,  which can be approximated by some effective screen   $f_{lens} [U_{in},D]$. Note that in relation \eqref{eq2} we both defined the effectiveg phase screen on a dense grid $\hat{D}_f [U_{in}]$ and introduced its approximation on a sparse grid $f_{lens}$, which we will call an effective lens (EL).

EL is specified as a nonlinear functional of the incoming field $U_{in}$ and several (yet unknown) set of the problem parameters $D$, which must be found based on minimizing the residual value of the exact numerical (true solution) and effective solution on a set of test examples.  In this case, the measure of the difference between the true solution and the effective one, i.e. the discrepancy, is determined to some extent arbitrarily based on the physical requirements of a specific problem being solved.
Generally, it is necessary to monitor the drop in the optical pulse energy as it propagates in a nonlinear dissipative medium, the structure of the filament and postfilament field surroundings, the dynamics of the maximum pulse intensity along the path, as well as the evolution of the NLSE Hamiltonian \cite{Bulygin_17}. Accordingly, when studying the properties of a post-filamentation pulse propagation, it is necessary to trace mainly its parameters, which determine the residual between the exact and approximation solutions.
The generation of such EL is based on the solution to the classical problem of a single laser pulse filamentation in a nonlinear medium using the method of the Ansatz \cite{Couairon} in combination with any ML methodic, e.g., the genetic algorithm (GA). GA allows one to search for a solution to the original dynamic problem for a field function on a fixed class of test functions provided that the resulting function must minimize the residual functional from the true solution. Here, by the true solution we mean the set of solutions obtained on a dense grid through the classical NLSE numerical solution.
To find a set of test solutions, it is necessary to solve the nonlinear Schrödinger equation for an optical pulse envelope propagating in a nonlinear medium. In this case, it is necessary to take into account not only the cubic (Kerr) nonlinearity, but also the physical effects associated with higher optical nonlinearities causing the beam transverse collapse arrest and the realization of pulse filamentation regime.
We consider the NLSE-type pulse propagation equation in the following form:

\begin{eqnarray}\label{eq3} 
&& {2ik_0 \partial_z  U}=\hat{h}_{k} U+ \nonumber\\ 
&&+ 
(\epsilon_m (UU^* )^{2m}+ik_0 \alpha_m (UU^* )^{2(m-1)})U		 
\end{eqnarray}

Here, $\epsilon_m$ and $\alpha_m$ are the coefficients for the m-th order nonlinearity leading to the beam self-focusing arrest. These coefficients account for the physical effects simulating plasma refraction and nonlinear absorption in self-induced plasma, respectively. In our analysis, the numerical solution to (\ref{eq3})), which provides the conservation of all integrals of motion, is implemented using a semi-implicit finite-difference scheme \cite{Balashov_2006,Bulygin_17}.
One of the main challenges for seeking the Ansatz for Eqs. (1-3) is that the optical field variation in the region of pulse self-compression near the nonlinear focus, where the higher-order nonlinearities strongly manifest themselves has characteristic transverse dimensions of about 100–200 mkm. At the same time, when leaving the collapse region, a pulse is localized on rather larger spatial scales of about 500 µm. Meanwhile, this region hosts the formation of a ring structures in pulse transverse profile, which is specific for laser filamentation (see Fig. la below). This means that the effective screen, which depends on the input field, must be the nonlocal functional of the input optical field.
As one of the possible options for constructing the desired EL we can propose a procedure, according to which the optical intensity (squared electric field module), $w=UU^*$  , is subjected by some kind of a diffraction operator upon reaching some chosen "pre-filamentation" value $w_{pf}$. In the following, this value is selected in the range from 2 $TW/cm^2$ to 8 $TW/cm^2$. In other words, if the condition $w =w _{pf}$ is satisfied, a new field  $w_{dif}$ is constructed as follows:
\begin{equation}\label{eq4} 
w_{dif}=(\hat{F}^{(-1)} exp(-l_{dif})\hat{F})w.		 
\end{equation}
Here, two undefined parameters are introduced, namely $w_{pf}$ and $l_{dif}$. In this case, the symbol$ \hat{F} $ in Eq. (4) denotes the Fourier transform operator. Importantly, the spatial region containing the field maximum has larger dimensions than the dimensions of the surroundings of the maximum of the original field $w$. At the same time, $w_{dif}$ monotonically decreases when moving away from the centers of its extremes.
Next, to define the EL we trim the field $w_{dif}$ by some boundary level $w_l$ and obtain the desired lens field profile $w_{lens}$ (see Fig. 1). The boundary value $w_l$ is a free parameter of the model and is chosen in such a way that after the trimming procedure the dimensions of the generated EL is about ten-times the typical filament diameter, i.e., about 1000 µm. As a result, one obtains the EL array as a set of disconnected regions within the beam field profile.

%\begin{figure}[h]
%	\includegraphics{/home/and/Desktop/2022/2022/paper_Geints_2/graph/g1.pdf}% Here is how to import EPS art
%	\caption{\label{fig:epsart} A figure caption. The figure captions are
%		automatically numbered.}
% \end{figure}

\begin{figure} [ht!]
	\begin{center}
		\begin{tabular}{c} %% tabular useful for creating an array of images 
			\includegraphics[scale=0.36]{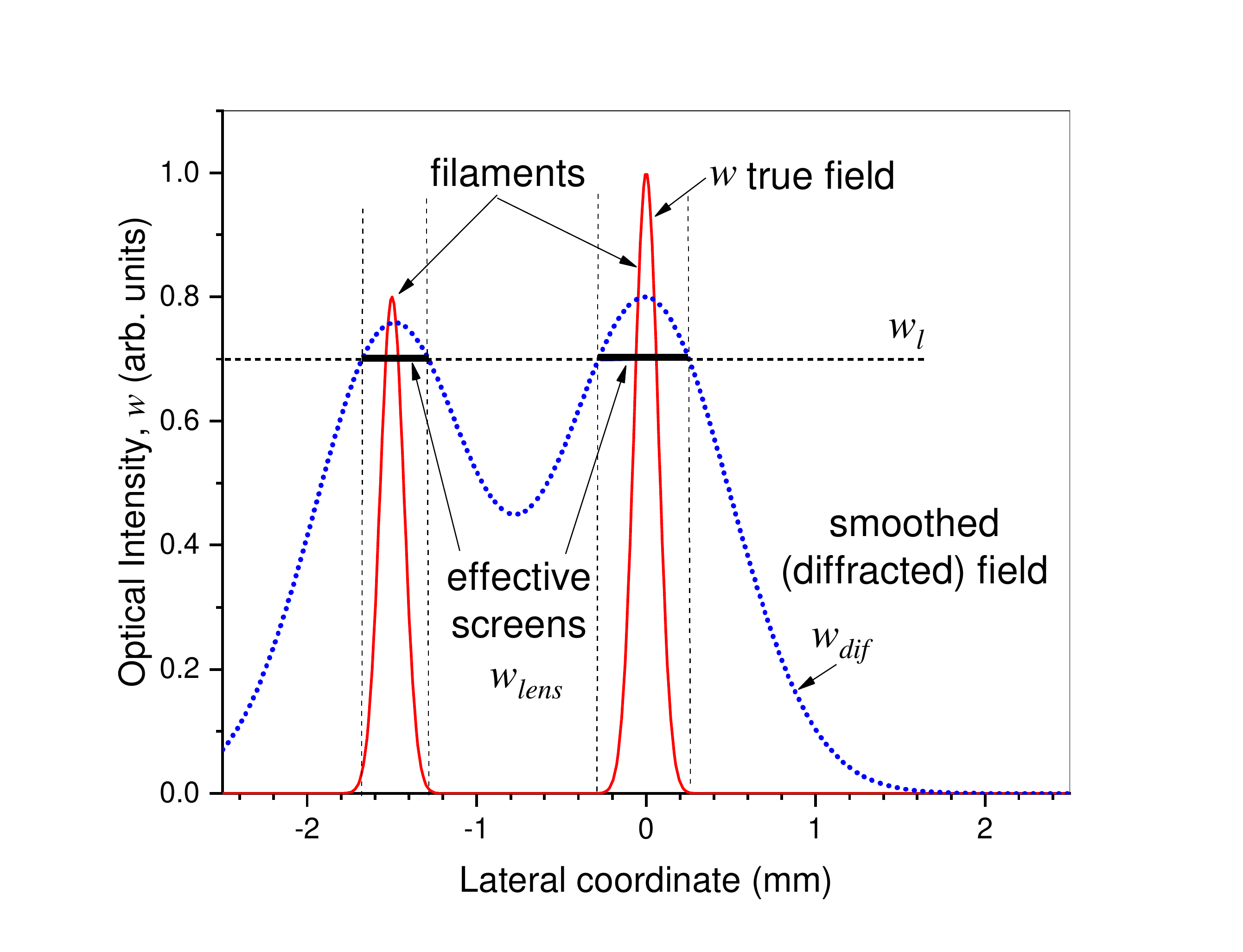}
		\end{tabular}
	\end{center}
	\vspace*{-10mm} \caption[example] 
	{ \label{fig 1} 
		Schematics of the procedure for EL generation for a given cut-off level $w_l$  }
\end{figure}

Due to the monotonicity of EL field $w_{lens}$, there is an unambiguous relationship between the distance from the EL centers and the value of the $w_{lens}$ field. This allows one using the inverse procedure to obtain the required lens fields $w_{lens}$, i.e.:
\begin{equation}\label{eq5} 
D_f [U_{in},d]= \sum_{r_c}f_{lens} (r-r_c)=f_{lens} (w_{lens}^{(-1)}).		
\end{equation}
Here, $r_c$ are the coordinates of EL center positions. In particular, based on the numerical simulation of single filamentation of submillimeter laser beams at the wavelength of 800 nm in air, the EL field profile $w_{lens}(r)$ can be approximated by the cubic dependence: $w_{lens}=c_1 (r^3-c_2)$ , where c1 and c2 are certain numerical coefficients depending on the pulse energy.
Consider the form of the chosen substitution function (Ansatz), which depends on the distance to the EL center obtained from the analysis of direct numerical simulation of pulse laser filamentation:
\begin{eqnarray}\label{eq6} 
&& f_{lens}=\sum_j a_j  cos(r\omega_j+\psi_j)(\theta(r_j))-\theta(r_{(j-1)}))+\nonumber\\
&&+
i\sum_l b_l  cos(r\nu_l+\eta_l)(\theta(r_l))-\theta(r_{(l-1)}))
\end{eqnarray}

Here, $\theta$ is the Heaviside function, and the remaining parameters define a set of free optimization parameters for the real $(r_j,a_j,\psi_j,\omega_j)$ and imaginary parts $(r_l,b_l,\eta_l,\nu_l)$ of the EL, respectively.
The requirement of sufficient EL smoothness (no discontinuity of the function and its derivatives at the joining points) imposes additional restrictions on the value of these functions. As a result, the vector $D$ (gene) used in the genetic algorithm has the dimension $(2n_g+3)$, where $n_g=3$ is the number of joining points. It should be noted that the best gene $D$ which is found by GA for one variant of the problem being solved may not be optimal for another problem statement. So, when varying the initial data of the test problem simulating a single pulse filamentation in air in terms of changing the pulse power, beam radius, focusing, etc., the best gene can have different form. However, this just means that one needs to expand the number of the control parameters by making it (selection) more flexible. 
The search for the optimal parameters of the Ansatz is a classical optimization problem belonging to the field of mathematical programming. Due to the fact that the problem considered is multi-parametric, it is instructive to solve it using the GA method. In this study, one of the variants of the classical GA implementations is chosen, which in particular can be implemented for the classical problem of building the optimal lens profile for optical field focusing with minimal wave aberrations \cite{Hoschel}.
For implementing the approach described above it is necessary to solve the reference (test) problem defined by Eq. (3) for single filamentation of an optical pulse by considering a set of initial beam profiles in the spatial form of a Gaussian beam $U_n$ with e-width $r_n$ and initial amplitude $A_n$ $(n = 1..N)$:
\begin{equation}\label{eq7}
U_n (r)=A_n exp(-(r/r_n )^2)
\end{equation}
The range of pulse initial power $P=(\pi A_n  r_n^2) $ and beam size used in the simulations is determined by the typical scales of field perturbations which seed the filaments in air, i.e. we chose P  from  3 Pcr to 9 Pcr and rn from 0.5 mm to 2.5 mm \cite{Talanov}, where Pcr is the critical self-focusing power of a Gaussian beam (Pcr  varies from 4 GW to about 7 GW in air in the near-IR range  \cite{COUAIRON_2007}. The next step required for EL building procedure is the determination of proper measure of the difference between the solutions generated using the EL method on a sparse grid and test solutions found on a dense grid. This will allow one to calculate the penalty function required for the evolutionary algorithm to work. Assume, that the solutions of problem (3) obtained by the effective lens method can be considered as acceptably close to the test ones according to the criterion defined through the difference in the spatial field moments:
\begin{equation}\label{eq8}
Q^{mn}=\int (U_n U_n^* )^2m ( x^2+y^2 )^m dxdy,	
\end{equation}		 
Recall, that another important issue of this measure is the length of the filamentation region and the rate of pulse intensity decrease on the beam axis, which reflects the dynamics of the post-filamentation pulse propagation.
The block diagram of GA implementation for searching the Ansatz is shown in Fig. 2. Here, the vector P characterizes the solution for the gene D. P is used for construction the so-called population fitness function f with the help of the Euclidean metric. In order to optimize the GA, the creation of the first viable generation (elite generation with Del gene) is carried out with 5 representatives by selecting from random implementations of generations. Population sizes vary between 5 and 10 chromosomes. The condition for the viability of a generation when finding an elite gene is the value of the deviation of the vector P from the vector P0, which in turn is found from the solution of the problem (3) on a dense grid for a set of sub-millimeter Gaussian beams.
In a random enumeration of genes, those of them are selected that according to the norm are at the smallest distances from P0 than the given measure fel. In its turn, fel value is selected in such a way that approximately one of 10 random representatives is viable. Thus, we get the first viable generation with a certain set of Del genes. Worthwhile, the set of Gaussian beams is specially chosen to produce a single laser pulse filamentation. However, if necessary, the proposed methodology allows generalizing this approach for the case of coupled multiple filaments.
\begin{figure} [ht!]
	\begin{center}
		\begin{tabular}{c} %% tabular useful for creating an array of images 
			\hspace*{-1cm}
			\includegraphics[scale=0.45]{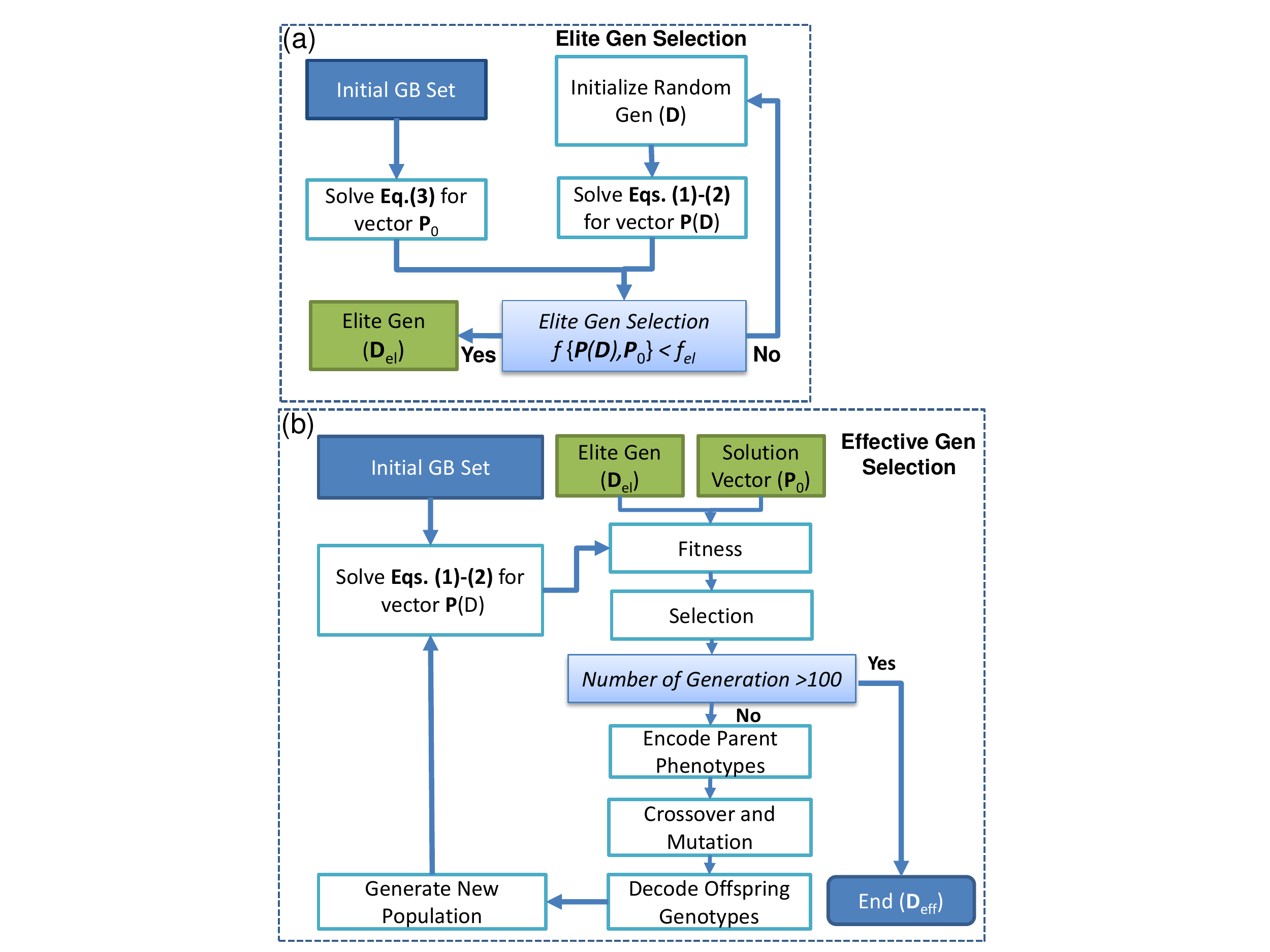}
		\end{tabular}
	\end{center}
	\vspace*{-5mm} \caption[example] 
	{ \label{fig 2} 
		Flow diagram for obtaining (a) elite gen and (b) effective gen for the efficient filamentation model based on a genetic algorithm.  }
\end{figure} 

The result of the GA operation is some efficient gene Deff, which provides the best set of Ansatz function parameters that completely define the effective screen. The selection is carried out by the roulette method. If the launched evolution goes in the wrong way, i.e. if the current state of the generation is worse than the elite implementation, then the destruction of degenerative descendants is carried out and the elite generation is reproduced again.
The longitudinal size of the EL in this implementation is fixed and equals to 30 cm, because it is this value that correlates with the build-up distance of the filament from the moments when pulse intensity reaches a pre-filamentation value $w_{pf}$ until exiting from the collapse at the same intensity value. At the same time, the behavior of the field in the vicinity of the beam collapse has a fairly universal form and weakly depends on the parameters of the incoming optical field \cite{Boyd,Talanov}.
Recall, the direct solution to the Eq. (3) requires a sufficiently dense numerical grid. Thus, the transverse grid step should less than 10 mkm (for pulse filamentation in air) and under the action of plasma and other higher nonlinearities the longitudinal step can decrease to micrometer values, which significantly complicates the implementation of massive numerical calculations. When using the ES approach in the filamentation simulation, the requirements for the grid step are reduced by at least an order of magnitude, and in the region of nonlinear effects manifestation the longitudinal step can be reduced by two or more orders of magnitude. This makes it possible to carry out large-scale numerical simulations even for centimeter-diameter beams propagating in real atmospheric conditions.
\section{EL benchmarking}
\subsection{EL benchmarking by a single pulse filamentation regime}
To demonstrate the efficiency of the EL technique, consider the case of a single fila-mentation in air of an unfocused laser pulse with radius of 1 mm and a power of 5 Pcr.
In Fig. 3, an illustration of the generated effective lens flens is presented, which is used to simulate the traversal of the first nonlinear focus by the laser beam. As noted above, this situation corresponds to the transverse collapse arrest due to the air plasma nonlinearity. In the figures below, the complex-valued optical thickness $D_f$ of the beam channel is shown by the open points, obtained by direct numerical solution of the problem (3) on dense spatial grid. The complex phase screen $D_f$ is built based on the Eq. (2) and the numerical solution of Eq. (3) on a dense numerical grid. The fields Uin and Uout are chosen at the first nonlinear focus. As noted above, the entry and exit conditions of the nonlinear focus correspond to the limiting value of the field fluence on beam axis equal to wpf. The solid curves in these plots are the approximations of EL dielectric constant flens based on the expression (5). 
The result of the energy density transformation of the optical field after passing the ef-fective lens can be seen in Fig. 4. In this figure, for comparison, the transverse distribution of the normalized fluence w(x,y) is given as two parts of one image combined along the vertical axis. Namely, on the left side of the image one can see the pulse fluence obtained as the exact solution to the stationary NLSE (3), and on the right the same profile is plotted but calculated through the EL model (6). As can be clearly seen, the beam profiles obtained by both methods are similar in their structure and exhibit even number of conical emission rings.
It is also important to compare the global characteristics of laser beam evolution (as a whole) along the optical path. Namely, we compare the peak intensity Im and the total pulse energy normalized to its initial value in the rigorous and effective propagation models. These parameters are shown in Figs. 5(a,d). As seen, the difference between the effective model and the test solution for pulses with different initial power  $P_0 $ is less than $15\% $  in terms of the total pulse energy. The most noticeable discrepancy is observed in the behavior of the maximum pulse intensity at the moment of filamentation beginning, which can achieve one order of magnitude. However, on average, despite the differences in the peak intensity dynamics the values of the filamentation length obtained by the EL method are reproduced with sufficient physical accuracy.

\begin{figure} [ht!]
	\begin{center}
		\begin{tabular}{c} %% tabular useful for creating an array of images 
			\hspace*{-0.5cm}
	\includegraphics[scale=0.2615]{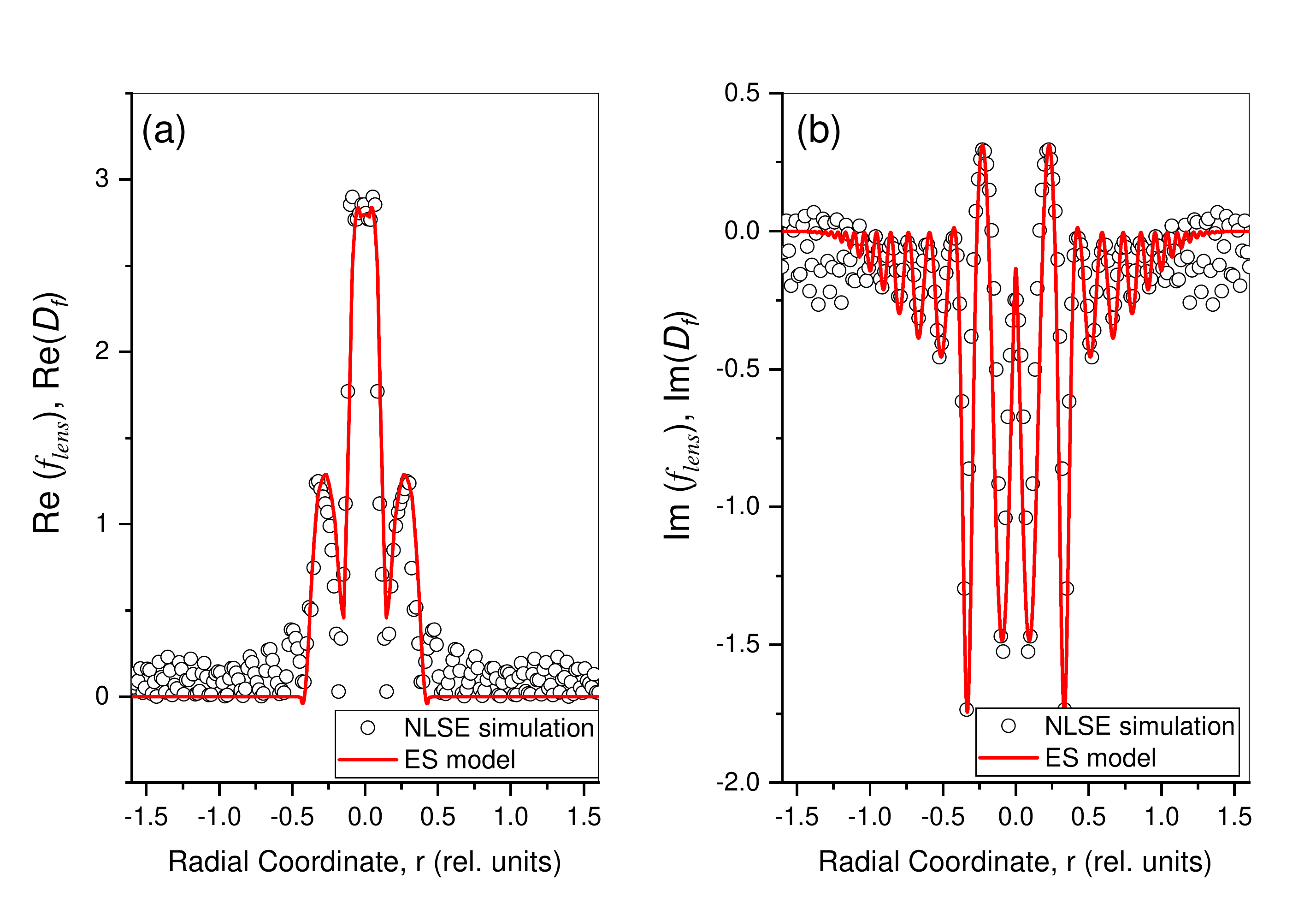}
	\end{tabular}
	\end{center} %	\vspace*{-5mm} 
	\vspace*{-5 mm} \caption[example] 
	{ \label{fig 3} 
		Radial distribution of (a) real and (b) imaginary parts of medium optical thickness $D_f$ (open points) and effective lens $f_{lens}$ (red curve) constructed for the case of single laser pulse filamentation in air (beam radius 1 mm, power 5 Pcr).   }
\end{figure}

\begin{figure} [ht!]
	\begin{center}
		\begin{tabular}{c} %% tabular useful for creating an array of images 
			\hspace*{-0.085cm}
	\includegraphics[scale=0.215]{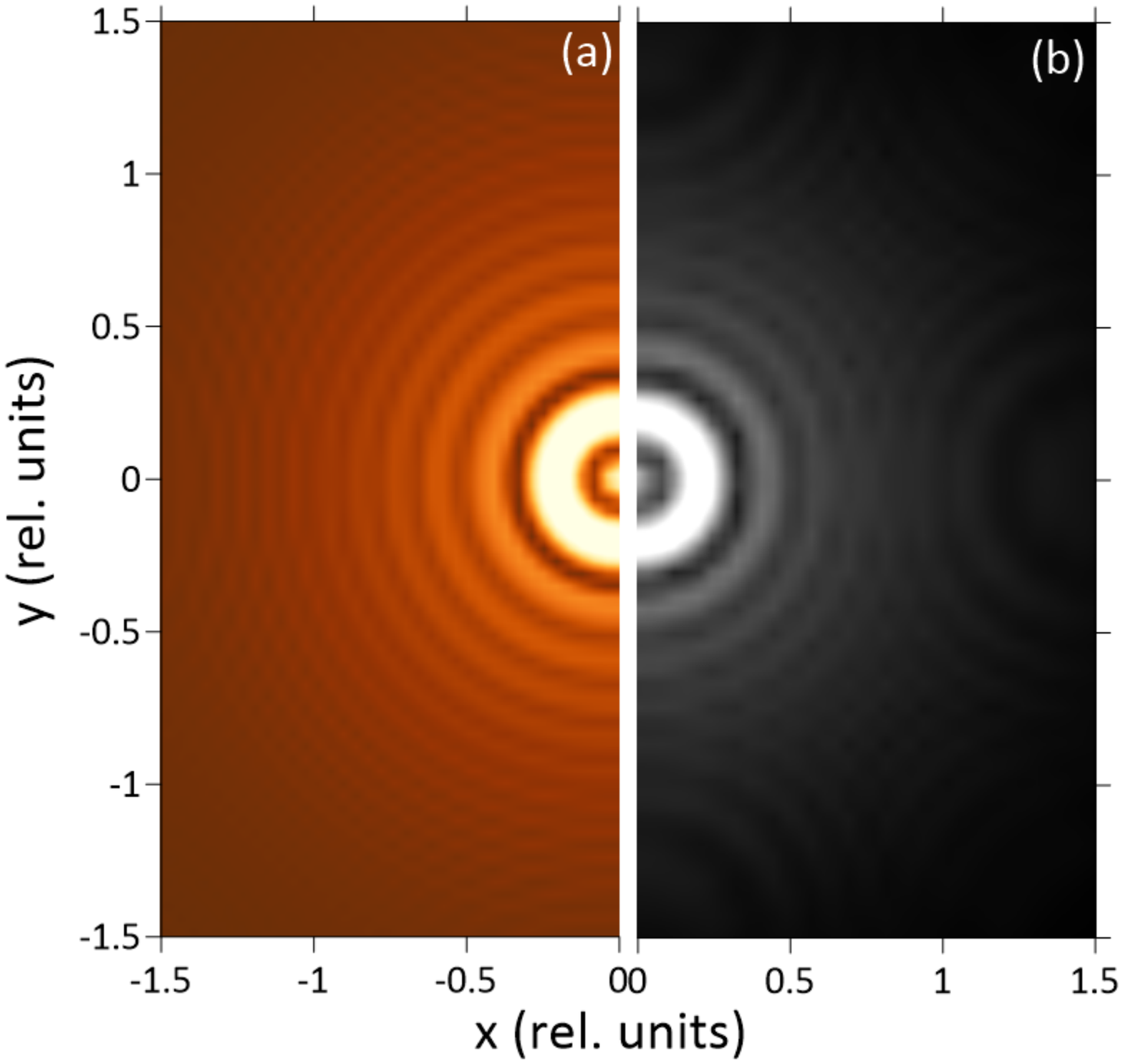}
	\end{tabular}
	\end{center} %	\vspace*{-5mm} 
	\vspace*{-5 mm}\caption[example] 
	{ \label{fig 4} 
		Normalized pulse fluence w(x,y) after passing the nonlinear focus. The left half-image (a) is obtained by the exact NLSE (3) solution, right half-image (b) represents the fluence profile obtained by EL method.    }
\end{figure}

\begin{figure} [ht!]
	\begin{center}
		\begin{tabular}{c} %% tabular useful for creating an array of images 
			\hspace*{-0.385cm}
			\includegraphics[scale=0.18415]{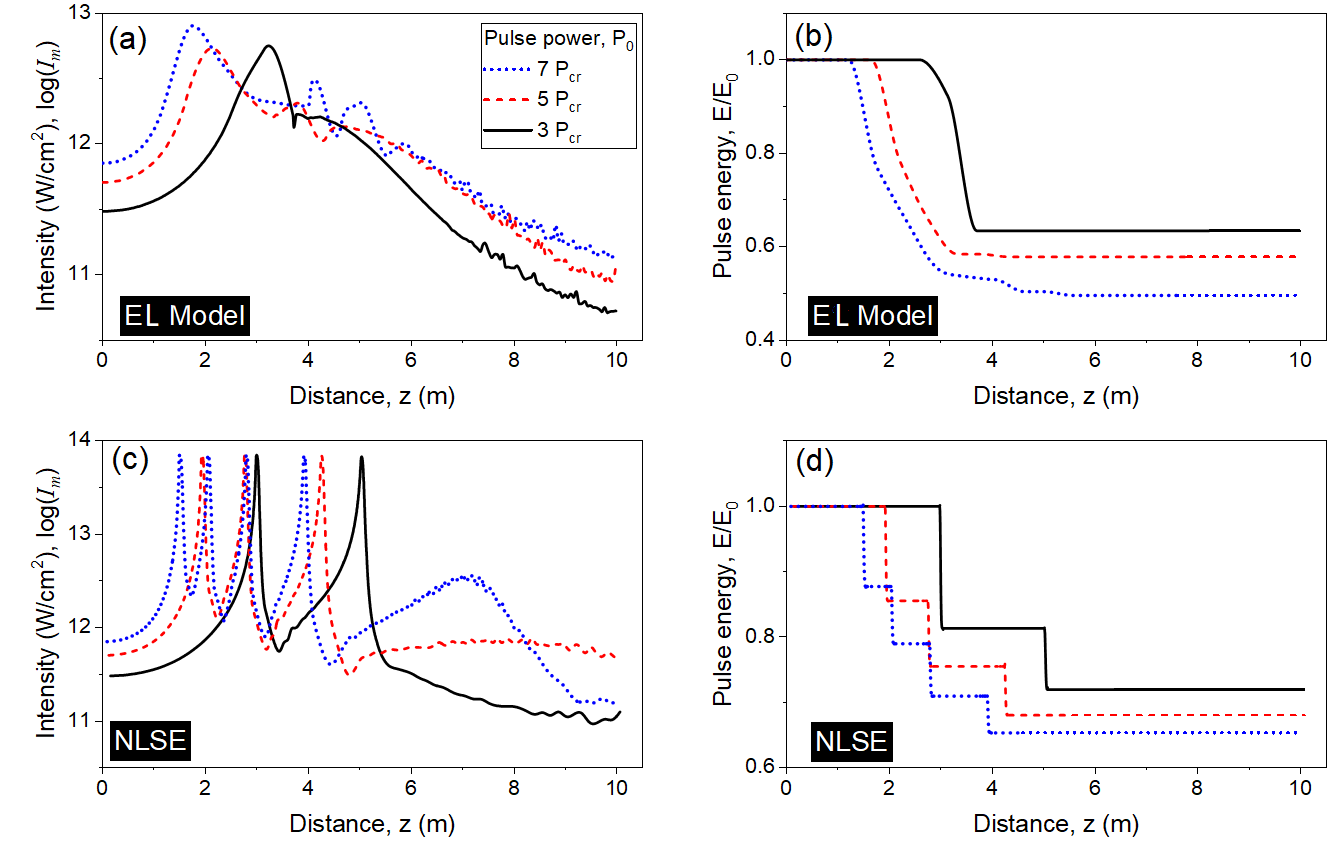}
			%%\includegra
		\end{tabular}
	\end{center} %	\vspace*{-5mm} 
	\vspace*{-5 mm} \caption[example] 
	{ \label{fig 5} 
		Normalized pulse fluence w(x,y) after passing the nonlinear focus. The left half-image (a) is obtained by the exact NLSE (3) solution, right half-image (b) represents the fluence profile obtained by EL method.    }
\end{figure}

\subsection{EL benchmarking by a single pulse filamentation regime}

In this Section, we present the results of the benchmarking of the proposed numerical method for the multiple filamentation mode. As an example, consider the initial Gaussian beam with two superposed intensive narrow side lobes, as shown in the inset to Fig. 6. These additional lobes serve as the seeds for off-axis filaments formation. We choose the initial beam diameter as 1 cm and set the pulse power as $P_0 = 380 GW = 112 Pcr$. In this case, it is important to analyze not only the change in pulse energy but also the dynamics of the number $N_{ha}$ of the most intense "hot areas" formed inside the beam. The filaments subsequently evolve into the so-called postfilament channels (postfilaments), which extend over a sufficiently long distance having subdiffraction angular divergence and moderate peak intensity (up to 0.1 $TW/cm^2$)  \cite{Bulygin_7,Geints_2007,Daigle,Geints16}. The condition,$ w \ge w_p$, where $w_p$ is certain selective cut-off intensity value, is used to distinguish "hot areas " inside the laser beam. The trace evolution of the normalized laser pulse energy and the $N_{ha}$ parameter are shown in Fig. 6(a) and (b), respectively.

\begin{figure} [ht!]
	\begin{center}
		\begin{tabular}{c} %% tabular useful for creating an array of images 
			\hspace*{-0.5cm}
	    	\includegraphics[scale=0.3015]{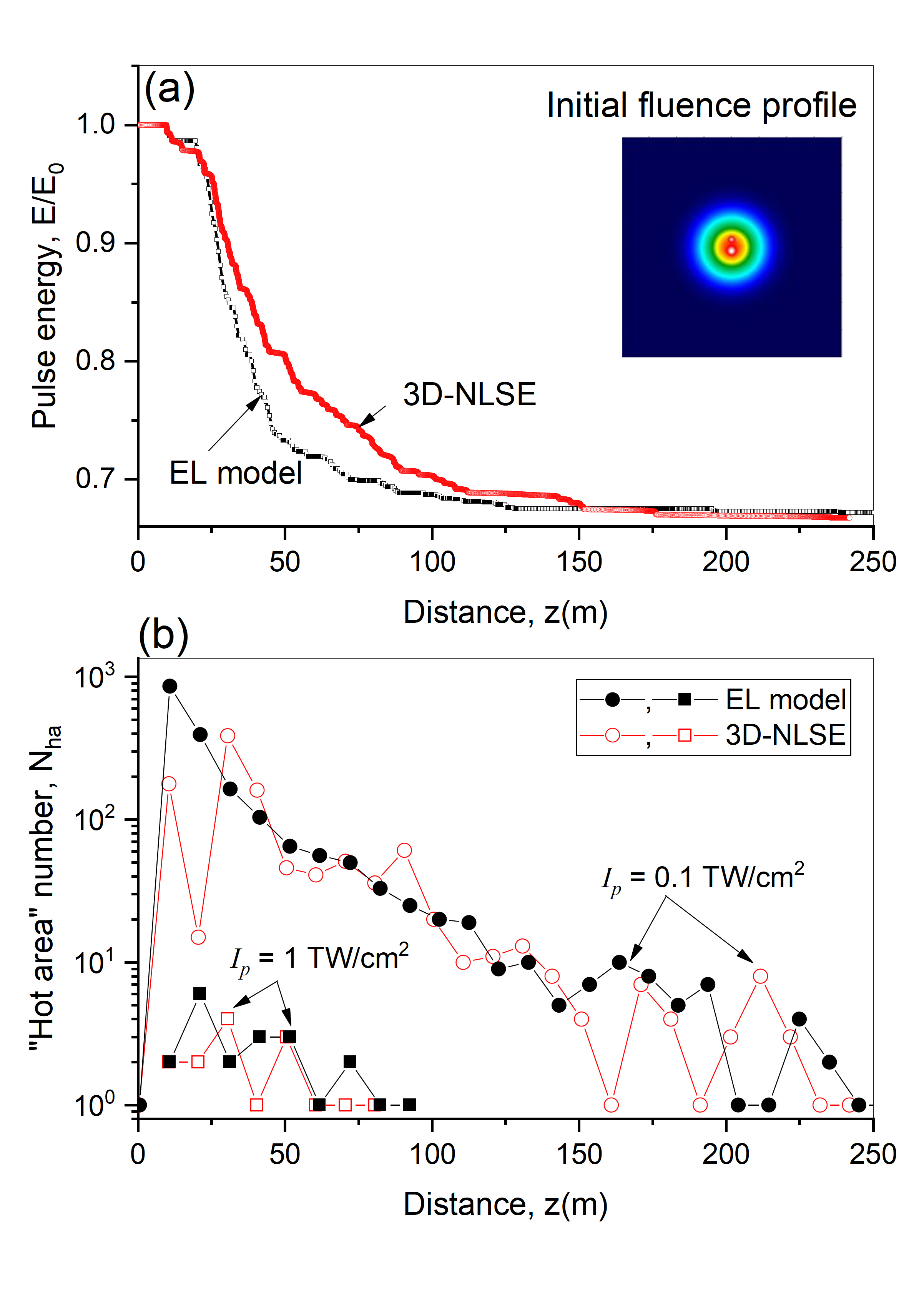}
	\end{tabular}
	\end{center}  	\vspace*{-10mm} \caption[example] 
	{ \label{fig 6} 
		Multiple filamentation of a 1 cm laser beam in air: Comparison of direct 3D-NLSE calculation and EL method. Variation along the optical path of (a) total pulse energy E (inset - initial beam profile) and (b) number of "hot areas" $N_{ha}$ at different intensity cut-off level $w_p$.  }
\end{figure}

Clearly, the figures show a satisfactory agreement between the two filamentation sim-ulation techniques. Furthermore, the correspondence is observed not only in terms of the qualitative behavior of the pulse parameters, but also in quantitative meaning. This validates the possibility of using EL method for the simulation the multiple filamentation of moderate-aperture laser beams under the conditions of multiple “hot areas” formation inside a beam.

\subsection{Multiple filamentation regime in turbulent air}
The above presented results show that the EL method is suitable for numerical simulation of single and multiple filamentation of a femtosecond laser pulse with beam diameter up to several centimeters in clear atmosphere. However, in real situations a wide-aperture femtosecond radiation often propagates in a turbulent air. In this section, we demonstrate the flexibility and robustness of the proposed EL numerical method for such situations. We give the results of our simulation of a sub-terawatt laser pulse filamentation with larger diameter (5 cm) on an airborne trace. On this air path we create an artificial turbulent layer in order to simulate the atmospheric turbulence.
Note that such numerical calculations are very challenging even within the stationary 3D-NLSE model, since this requires hundreds of CPU-hours and huge computing resources by using the modern supercomputers. Alternately, the proposed EL method gives a significant increase in computational performance. For example, the average per task runtime (statistical realization) of the considered 3D problem takes about 15 to 20 minutes on a personal computer with average computing performance. 
To test the efficiency of the proposed numerical technique, we carry out a numerical experiment on the filamentation of high-power pulse of titanium-sapphire femtosecond laser (peak pulse power up to 0.5 TW) on a long air path ($\sim 100 m$ ) with an optical turbulence. We numerically reproduce the experimental condition previously reported in Ref. [28], where one can find detailed information on the methodology of these experiments. In the simulations, air turbulence is set in the form of a spatially localized (about a meter wide) turbulent phase screen which can be placed in different parts of the optical path. Note, in real experiments this turbulence layer was created using an industrial air heater. The structural composition and the strength of the turbulence were controlled by changing the temperature of the air jet. As shown in \cite{Apeksimov22}, chaotic modulations of the laser beam supported by the optical nonlinearity of the medium led to the small-scale self-focusing of laser pulse, its spatial fragmentation and the appearance of multiple high-intensity light spots ("hot areas") along the propagation path. Unlike \cite{Apeksimov22}, here we consider different yet unpublished results of these experiments obtained for a laser beam with larger size of 5 cm (not 2.5 cm as in \cite{Apeksimov22}), which is produced by telescoping the initial narrow beam emitted from the femtosecond oscillator output.
Recall, that experimentally, the turbulent air layer was created in the form of a hot jet with stepwise variable temperature T at the outlet of the fan nozzle from 100 to 600 $C^o$. According to our estimation, the values of structural parameter   of the artificial turbulent layer are much higher than the typical values of atmospheric turbulence and exceed  $\sim 10^{-10} m^{-2/3}$. In accordance with real experiments, it is also necessary to numerically reproduce the presence of a turbulent air layer on the optical path.
To this end, the NLSE (3) is modified by including along with the effective complex lens $f_{lens}$ also a random turbulent phase screen $\epsilon_t$:

\begin{equation}\label{eq9} 
2ik_0 \partial_z  U=\hat{h}_{k} U+ (f_{lens}[U,D]+\epsilon_t)U		 
\end{equation}

The turbulent phase screen is constructed in a regular way by the spectral method [26]. In this case, to simplify the calculations the spectral density function of turbulent inhomogeneities $\epsilon_t$: is modeled by a step-function:
\begin{equation}\label{eq10}							
\epsilon_t=\tau \theta(k-k_{h})
\end{equation}
Here, $k_{h}$ is the upper cut-off spatial frequency of the pulse spectrum. The spectral amplitude of turbulent perturbances $\tau$ is associated with the tem-perature of the air heater as $T=\nu \tau k_{h}^2$ (here $\nu$ is an fitting parameter).

Figs. 7(a, b) shows the constructed turbulent phase screens (in normalized variables) for two different temperatures T of the air jet as an example. The corresponding results of EL method application for simulation laser beam filamentation in turbulent air are presented in Figs. 7(c, d). In the simulations, the pulse energy is set to 35 mJ (peak power ~400 GW), and the distributions of the normalized fluence in the beam cross-section are given at the spatial coordinates right before and inside a multiple filamentation region in air.

\begin{figure} [ht!]
	\begin{center}
		\begin{tabular}{c} %% tabular useful for creating an array of images 
			\hspace*{-0.5cm}
			\includegraphics[scale=0.3615]{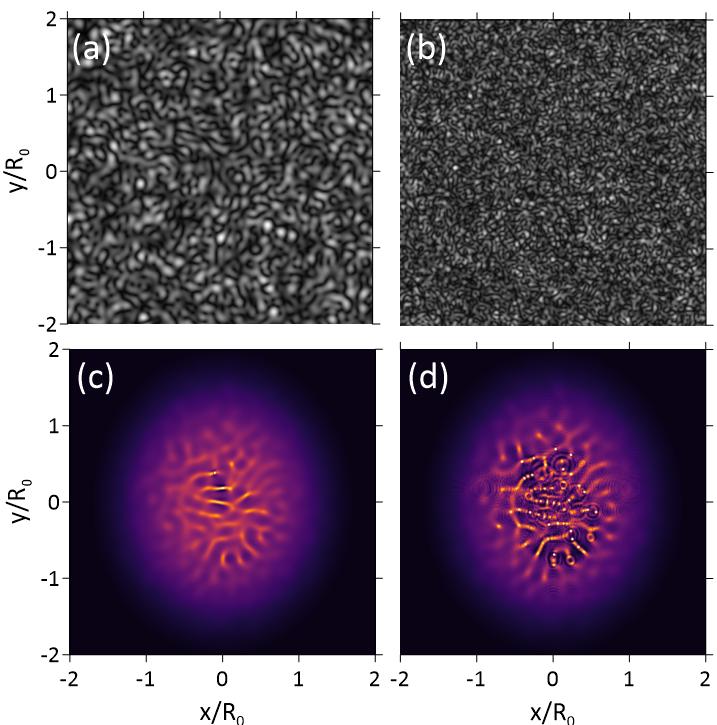}%%\includegra
		\end{tabular}
	\end{center}  	\vspace*{-5mm} \caption[example] 
	{ \label{fig 7} 
		(a, b) Examples of turbulent phase screens used in multiple filamentation simulations for T = 100 $C^o$ (a) and 400 $C^o$(b). (c, d) Simulated spatial distribution of normalized fluence $w/w_0$ (on a logarithmic scale) at 25 m (c) and 35 m (d).   }
\end{figure}

As seen from these distributions, the proposed EL method numerically reproduces the typical pattern of multiple filamentation of the beam \cite{Henin}. The induced thermal source in-homogeneities of the optical field are clearly distinguishably in the laser beam in Fig. 7(c), which seed later the filaments visible in Fig. 7(d) as the "hot areas".
To be more specific, we characterize the "hot areas" by certain threshold pulse intensity value, say $w_{p}$ = 0.01 $Tw/cm^2$. Consequently, a "hot area" is defined as a closed region, where $ w \ge w_p$. Then, we calculate the "hot area" number $N_{ha}$ by counting the closed regions in the laser beam at optical path end and present the results in Figs. 8(a) and (b).

\begin{figure} [h!]
	\begin{center}
		\begin{tabular}{c} %% tabular useful for creating an array of images 
			\hspace*{-0.5cm}
			\includegraphics[scale=0.3115]{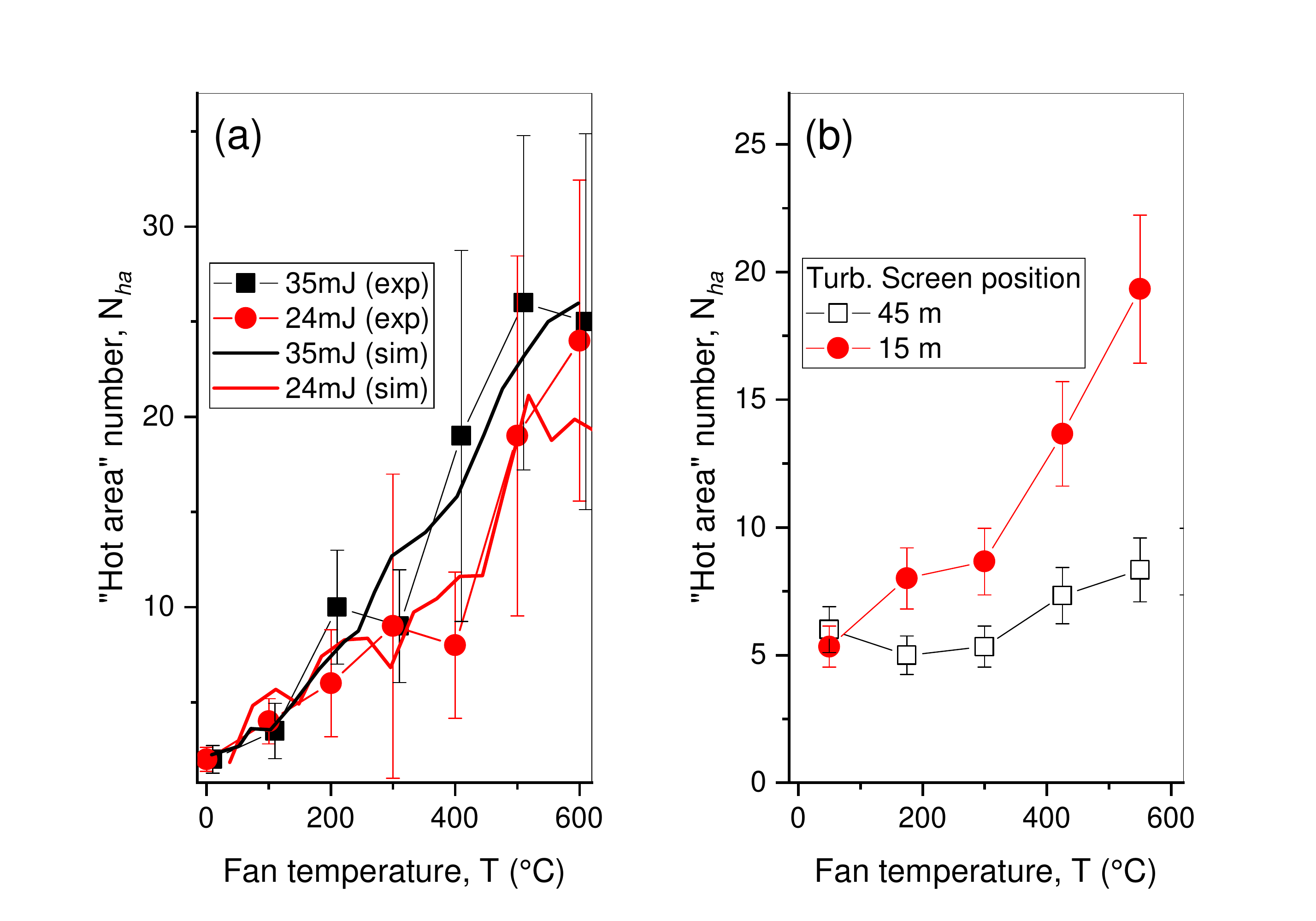}%%\includegra
		\end{tabular}
	\end{center}  	\vspace*{-5mm} \caption[example] 
	{ \label{fig 8} 
		(a) Comparison of the experimental (exp) and calculated (sim) number $N_{ha}$ of "hot areas" formed in a Gaussian laser beam with diameter of 5 cm and different energy when changing heater fan temperature T. (b) Simulated number $N_{ha}$ of "hot areas" at different location of the turbulent layer and pulse energy 35 mJ.  }
\end{figure}

\section{Discussion}

From Fig. 8(a) one can see, that the parameter $N_{ha}$ grows with increasing air jet temperature T. This trend is consistent with the change in the scale of inhomogeneities of the random phase screen (compare Figs. 7(a) and (b)) within the framework of the proposed theoretical model. Generally, this increasing dynamic is non monotonic. In this case, the interval of continuous$N_{ha}$ growth can be qualitatively explained by the Bespalov-Talanov modulation instability theory (BPT) \cite{Talanov}. BPT predicts that the increase in the perturbation of the optical wave phase leads to an increase in the number of "hot areas" in the wave amplitude profile when propagating in a Kerr-type medium. However, only those perturbations will be supported by the self-focusing optical nonlinearity, in which the nonlinearity suppresses the diffraction, i.e. which carry sufficient optical power larger than the critical value Pcr. In this case, the greater the average intensity along the beam profile, the smaller the scale of viable perturbations. In our theoretical model, an increase in fan temperature T causes greater fragmentation of the beam profile. Consequently, the number of arising "hot areas" is also growing.
It is important to emphasize that the above reasoning can only be regarded as a qualitative analysis of the dynamics of beam fragmentation in a turbulent medium, since the optical intensity is initially spatially inhomogeneous, decreasing in amplitude from the center to the beam periphery (Gaussian shape). In addition, the application of the BPT is correct only for small initial perturbations. In this regard, BPT predicts only the proportionality of the number of viable perturbations to their relative amplitude. However, the very type of this dependence seems to be nonlinear due to the diffraction coupling between different perturbations in the Kerr medium.
We calculate the number of "hot areas" formed in the laser beam when the turbulent layer is placed at different spatial positions along the optical propagation path. Specifically, the heater fan is placed at two positions: (a) in front of the pulse filamentation region, which corresponds to a propagation distance of 15 m, and (b) inside the filamentation region at a distance of 45 m from the laser source. The results of these numerical experiments for a 35 mJ pulse are shown in Fig. 8(b). Remember, the initial beam diameter is 5 cm.
From this figure one can see, that if the turbulence is positioned before the filamentation region the induced optical phase perturbations cause the monotonic growth of "hot area" number with the fan temperature. Obviously, at the beginning of multiple filamentation, a laser beam as a whole is more concentrated in space and has more energy than after the filamentation region, therefore, the optical radiation is more sensitive to any spatial per-turbations imposed by the turbulent layer.
Alternatively, by placing the turbulent screen in the region of "developed" pulse fila-mentation (open symbols), there is a weak growth and sometimes even a decrease of $N_{ha}$ (T < 200 $C^o$) with the increase of fan temperature. Similar tendency was reported earlier in similar experiments with a smaller beam size \cite{Apeksimov22,Houard08} . This demonstrates the impressive stability of light filaments and postfilaments to stochastic pulse phase perturbations under conditions of constant maintenance of the dynamic balance between the focusing Kerr and defocusing plasma nonlinearities of the optical medium. A qualitative explanation for this dependence can be found in the fact that in the region of strong nonlinearity, induced fluc-tuations cannot destroy the already formed inhomogeneities in the beam during it self-focusing, and for a low-intensity background they almost no longer affect the conditions for forming new "hot areas". Indeed, when a turbulent screen is placed in a region where filaments have already been formed, its role in the formation of "hot areas " is significantly reduced, since the optical path must be of sufficient length to convert the corresponding phase distortions into new viable amplitude perturbations. Additionally, the phase-induced amplitude perturbations must receive enough optical power for their development, whereas to the end of the optical trace the pulse power density on average decreases.

\section{Conclusions}

In conclusion, we propose an effective numerical method for the simulation of high-power wide-aperture laser radiation propagation in a turbid nonlinear medium based on the 3D NLSE numerical solution. This method relays on replacing the regions with strong medium optical nonlinearity and, accordingly, the regions demanding fine numerical grid, by a number of complex-valued (effective) phase screens, which provide the scattering and absorption of optical radiation in a certain manner. The choice of the specific lens structure is based on the selection of the Ansatz (substitution function) to the NLSE with the parameters providing the best accordance to the test problem solutions. This Ansatz is searched by the ML methods (the genetic algorithm).
So far, in present form the proposed algorithm is implemented for a simple scenario of stationary laser beam filamentation (single and multiple) and does account for any nonsta-tionary physical effects (pulse group velocity dispersion, optical "shock" waves, retarded Kerr rotational effect, etc.), as well as the mechanisms occurring during filament stochastic clustering ("optical pillars" \cite{Skupin08}). At the same time, with this method we are able to reproduce quantitatively the experimental dependencies for the evolution of "hot areas" (postfilaments) number formed in a wide laser beam (5 cm in diameter) during the propagation on an extended air path with localized turbulence layer.
Worthwhile, the proposed method possesses great opportunities of the implementation both for the non-stationary case of multiple filamentation and for the possible accounting of filament clustering during the propagation of wide-aperture laser radiation. Basically, the developed numerical method for solving nonlinear partial differential equations of the parabolic type (NLSE, unidirectional Maxwell equation) is not limited to the problem of propagation of high-power femtosecond laser radiation. In this regard, the presented method has a broader significance, because it can be extended to other practical problems, such as vision problems in lossy turbulent media, optical energy transmission in a disperse aerosol medium, electromagnetic wave propagation in plasma, etc. It is also possible to use the machine learning methods and build a base of test solutions using the methodology we have proposed.

\begin{backmatter}

%	\bmsection{Acknowledgments}
% The authors acknowledge the colleagues from V.E. Zuev Institute of Atmospheric Optics A.V. Petrov,  D.V. Apeksimov and A.M. Kabanov for sharing the experimental data. The effective lens method is developed within the scientific program of the Ministry of Science and Higher Education of the Russian Federation. All method benchmarking is supported by the Russian Science Foundation grant (21-12-00109).  Results presented in Sections "Math and Equations" (GA) was carried out with the support of a grant under the Decree of the Government of the Russian Federation No. 220 of 09 April 2010 (Agreement No. 075-15-2021-615 of 04 June 2021)' 
  
%	\bmsection{Conflict of interest}

%	The author declares no conflicts of interest.
		
\end{backmatter}

%%%%%%%%%%%%%%%%%%%%%%% References %%%%%%%%%%%%%%%%%%%%%%%%%

%%%%%%%%%% If using BibTeX:
 \bibliography{eff_l}

%%%%%%%%%% If preparing manually:
% \begin{thebibliography}{1}
% \newcommand{\enquote}[1]{``#1''}

% \bibitem{Zhang:14}
% Y.~Zhang, S.~Qiao, L.~Sun, Q.~W. Shi, W.~Huang, L.~Li, and Z.~Yang,
%   \enquote{Photoinduced active terahertz metamaterials with nanostructured
%   vanadium dioxide film deposited by sol-gel method,}
%   {\protect\JournalTitle{Optics Express}} \textbf{22}, 11070--11078 (2014).

% \bibitem{Optica}
% {Optica}, \enquote{{Optica Publishing Group},}
%   \url{http://www.opg.optica.org}.

% \bibitem{FORSTER2007}
% P.~Forster, V.~Ramaswamy, P.~Artaxo, T.~Bernsten, R.~Betts, D.~Fahey,
%   J.~Haywood, J.~Lean, D.~Lowe, G.~Myhre, J.~Nganga, R.~Prinn, G.~Raga,
%   M.~Schulz, and R.~V. Dorland, \enquote{Changes in atmospheric consituents and
%   in radiative forcing,} in \enquote{Climate Change 2007: The Physical Science
%   Basis. Contribution of Working Group 1 to the Fourth assesment report of
%   Intergovernmental Panel on Climate Change,}  S.~Solomon, D.~Qin, M.~Manning,
%   Z.~Chen, M.~Marquis, K.~B. Averyt, M.~Tignor, and H.~L. Miler, eds.
%   (Cambridge University Press, 2007).

% \end{thebibliography}

\end{document}